\documentclass{article}

\usepackage{microtype}
\usepackage{graphicx}
\usepackage{booktabs}

\usepackage{listings}
\usepackage{lipsum}

\usepackage{appendix}

\usepackage{hyperref}
\usepackage{subfig}

\newcommand{\smallsec}[1]{\vspace{-3mm} \paragraph{#1}}
\newcommand{\datasetname}{CoqGym}
\newcommand{\numberoffiles}{3,061 }
\newcommand{\numberofprojects}{123 }
\newcommand{\numberofproofs}{70,856 }
\newcommand{\numberoftrainproofs}{43,844 }
\newcommand{\numberofvalidproofs}{13,875 }
\newcommand{\numberoftestproofs}{13,137 }

\newcommand{\numberofhumantrainvalidsteps}{190K }

\usepackage[accepted]{icml2019}
\icmltitlerunning{Learning to Prove Theorems via Interacting with Proof Assistants}

\begin{document}

\twocolumn[
\icmltitle{Learning to Prove Theorems via Interacting with Proof Assistants}

\icmlsetsymbol{equal}{*}

\begin{icmlauthorlist}
\icmlauthor{Kaiyu Yang}{pu}
\icmlauthor{Jia Deng}{pu}
\end{icmlauthorlist}

\icmlaffiliation{pu}{Department of Computer Science, Princeton University}

\icmlcorrespondingauthor{Kaiyu Yang}{kaiyuy@cs.princeton.edu}
\icmlcorrespondingauthor{Jia Deng}{jiadeng@cs.princeton.edu}

\icmlkeywords{Automated Theorem Proving, Machine Learning, Benchmark, Dataset, Learning Environment, Interactive Theorem Proving, Coq, \datasetname, ASTactic}

\vskip 0.3in
]

\printAffiliationsAndNotice{}

\begin{abstract}
Humans prove theorems by relying on substantial high-level reasoning and problem-specific insights.
Proof assistants offer a formalism that resembles human mathematical reasoning, 
representing theorems in higher-order logic and proofs as high-level tactics.
However, human experts have to construct proofs manually by entering tactics into the proof assistant.
In this paper, we study the problem of using machine learning to automate the interaction with proof assistants.
We construct \emph{\datasetname}, a large-scale dataset and learning environment containing 71K human-written proofs from \numberofprojects projects developed with the Coq proof assistant. We develop \emph{ASTactic}, a deep learning-based model that generates tactics as programs in the form of abstract syntax trees (ASTs). Experiments show that ASTactic trained on {\datasetname } can generate effective tactics and can be used to prove new theorems not previously provable by automated methods. Code is available at \url{https://github.com/princeton-vl/\datasetname}.
\end{abstract}

\section{Introduction}
Given the statement of a theorem, simply push a button, and the proof comes out.
If this fantasy of automated theorem proving (ATP) were true, it would impact formal mathematics~\citep{mccune1997solution}, software verification~\citep{darvas2005theorem}, and hardware design~\citep{kern1999formal}.
In reality, however, state-of-the-art theorem provers are still far behind human experts on efficiently
constructing proofs in large-scale formal systems.

Consider this theorem: $0 + 1 + 2 + \cdots + n = \frac{n(n+1)}{2}$. 
As humans, we decide to prove by induction on $n$.
After solving the trivial case ($n = 0$), we complete the proof by applying the induction hypothesis and simplifying the resulting formula.
Simple as it is, the proof requires understanding the concepts (natural numbers, addition), mastering the proof techniques (induction), as well as problem-specific insights to drive the decisions we made.

What a typical theorem prover does, however, is to prove by resolution refutation:
It converts the premises and the negation of the theorem into first-order clauses in conjunctive normal form (CNF).
It then keeps generating new clauses by applying the resolution rule until an empty clause emerges,
yielding a proof consisting of a long sequence of CNFs and resolutions. While this provides a universal procedure, the CNF representation of simple formulas can be long, complicated and inscrutable, making it difficult to benefit from the higher-level abstraction and manipulation that is common to human mathematical reasoning. 

To work around the difficulties of ATP in practical applications, interactive theorem proving (ITP)~\citep{harrison2014history} incorporates humans in the loop.
In ITP, human users define mathematical objects formally and prove theorems semi-automatically by entering a sequence of commands called tactics.
The tactics capture high-level proof techniques such as induction, 
leaving low-level details to the software referred to as proof assistants. A successful sequence of tactics is essentially a proof written in the language of the proof assistant. It can also be viewed as a program that is executed by the proof assistant. 

ITP relies on humans in the loop, which is labor-intensive and hinders its wider adoption. However, at the same time, it is a blessing in disguise, in that it opens up a route to full automation---human experts have written a large amount of ITP code, which provides an opportunity to develop machine learning systems to imitate humans for interacting with the proof assistant. Indeed, some recent efforts have attempted to learn to generate tactics from human-written proofs and have obtained promising results~\cite{gransden2015sepia,gauthier2018learning,bansal2019holist}. 

However, existing approaches to the ``auto-ITP'' problem suffer from two limitations. One is the lack of a large-scale dataset. Prior work was trained and evaluated on no more than a few thousands of theorems~\cite{gransden2015sepia,gauthier2018learning,huang2018gamepad}, likely insufficient for data-hungry approaches such as deep learning. The other is the limited flexibility of the learned models in generating tactics. A tactic can be a sophisticated line of code with functions, arguments, and compound expressions, and the space of possible tactics is infinite. Prior work has limited flexibility because they generate tactics by copying from a fixed, predetermined set~\cite{gransden2015sepia,gauthier2018learning,huang2018gamepad,bansal2019holist}, and are thus unable to generate out-of-vocabulary tactics unseen in the training data. 

In this work we address these limitations by making two contributions: a large-scale dataset and a new method for tactic generation that is more flexible and adaptive.

\smallsec{\datasetname: A large-scale ITP dataset and learning environment}
We construct \emph{\datasetname}, a dataset and learning environment for theorem proving in proof assistants. It includes 71K human-written proofs from \numberofprojects open-source software projects in the Coq proof assistant~\citep{barras1997coq}, covering a broad spectrum of application domains, including mathematics, computer hardware, programming languages, etc.
Our dataset is much larger and more diverse than existing datasets, which consist of only a few thousands of theorems and cover only a handful of domains such as Peano arithmetic~\citep{dixon2003isaplanner} or the Feit--Thompson theorem~\citep{gonthier2013machine}. The scale and diversity of our dataset facilitate training machine learning models and the evaluating cross-domain generalization.

The learning environment of {\datasetname } is designed for training and evaluating auto-ITP agents.
The agent starts with a set of premises and a goal (theorem) to prove; it interacts with the proof assistant by issuing a sequence of tactics. The proof assistant executes each tactic and reports the results back in terms of a set of new goals. The agent succeeds when no more goals exist.

To make the challenging task more amenable to learning, we augment {\datasetname } with shorter proofs. They are synthesized from the intermediate steps of the original human-written proofs and may serve as additional training data.

\smallsec{ASTactic: A new method for tactic generation}
We develop \emph{ASTactic}, a deep learning model for tactic generation that is more flexible and adaptive than prior work. It generates tactics as programs by composing abstract syntax trees (ASTs) in a predefined grammar using the tokens available at runtime. To our knowledge, this is the first time learning-based AST generation is applied in the context of interactive theorem proving. 

ASTactic takes in a goal and a set of premises expressed as terms in Coq's language.
It outputs a tactic as an AST in a subset of Coq's tactic language.

Experimental results on {\datasetname } show that ASTactic can generate effective tactics. It can successfully prove 12.2\% of the theorems in the test set, significantly outperforming the built-in automated tactics (\texttt{auto}, \texttt{intuition}, \texttt{easy}, etc.) in Coq (4.9\%). More importantly, our model can be combined with state-of-art ATP systems~\cite{de2008z3,BCD+11, kovacs2013first, schulz2013system} and boost the success rate further to 30.0\%, which shows that our model has learned effective higher level tactics and has proved new theorems not previously provable by automatic systems.

\smallsec{Contributions}
In summary, our contributions are two-fold. First, we build \datasetname---a large-scale dataset and learning environment for theorem proving via interacting with a proof assistant. Second, we develop ASTactic, a deep learning model that learns to generate tactics as abstract syntax trees and can be used to prove new theorems beyond the reach of previous automatic provers.

\section{Related Work}
\vspace{1mm}

\smallsec{Automated theorem proving} Modern theorem provers~\citep{kovacs2013first,schulz2013system} represent theorems in first-order logic and search for proofs in resolution-based proof calculi.
The proof search has been significantly improved by machine learning~\citep{irving2016deepmath,wang2017premise,urban2011malecop,bridge2014machine,loos2017deep,kaliszyk2018reinforcement,rocktaschel2017end}.
However, it remains a challenging task due to the large search space; 
as a result, state-of-the-art provers do not scale to large problems.
In contrast to traditional first-order provers, we focus on theorem proving in the Coq proof assistant, which represents theorems and manipulates proofs at a higher level, offering the unique opportunity of learning from human proofs.

Some proof assistants allow a user to use existing ATP systems directly. For example, Sledgehammer~\citep{paulson2010three} translates theorems in the Isabelle proof assistant~\citep{paulson1994isabelle} to first-order logic.
It then proves the theorems using external provers and converts the proof back to Isabelle's tactics.
Similar ``hammers'' were developed for other proof assistants as well~\citep{kaliszyk2014learning, urban2004mptp, czajka2018hammer}.
The hammer-based approach essentially bypasses the proof assistant and outsources the work to external ATPs. 
In contrast, our model learns to prove theorems within the proof assistant using native tactics without hammers.

\smallsec{Learning to interact with proof assistants} There have been relatively few works in learning to interact with proof assistants. SEPIA~\citep{gransden2015sepia} learns from existing Coq proofs a finite-state automaton, where each transition corresponds to a tactic, and each path corresponds to a sequence of tactics. During test time, it samples tactic sequences defined by this automaton. Note that SEPIA can only choose from a finite set of tactics, and it tries the same tactics regardless of the theorem. That is, it will apply the tactic ``\texttt{apply x}'' even when \texttt{x} is not a valid term in the current context. In contrast, our model can generate an infinite number of tactics tailored to the current context.

TacticToe~\citep{gauthier2018learning} generates tactics by retrieving from the training data a small number of candidate tactics that have been used for theorems similar to the current goal. Each candidate is treated as a possible action and evaluated by a learned value function and by Monte Carlo Tree Search. Although more adaptive than SEPIA, the generated tactics are still chosen from a fixed set with predetermined arguments and may have difficulty generalizing to new domains with out-of-vocabulary terms.  

FastSMT~\citep{balunovic2018learning} generates tactics to interact
with the Z3 SMT solver~\cite{de2008z3}, which determines the satisfiability of a certain class of logical formulas. 
Compared to our model, FastSMT uses substantially simpler tactics---all of them have only boolean and integer arguments, whereas tactics in Coq can have compound expressions as arguments. As a result, the approach of FastSMT does not output ASTs and is not directly applicable to our setting.

\smallsec{Datasets for theorem proving}
Our work is related to many previous theorem proving datasets~\cite{sutcliffe2009tptp, bancerek2015mizar,gransden2015sepia,gauthier2018learning,huang2018gamepad}. Our work differs from prior work in that we focus on theorems in higher-order logic and proofs consisting of high-level tactics as opposed to first-order logic and low-level proofs, and we aim for larger scale and more diverse domains.

The closest prior work is GamePad~\citep{huang2018gamepad}. GamePad includes a tool for interacting with Coq as well as data collected from the proof of the Feit– Thompson theorem~\citep{gonthier2013machine}. We have independently developed a similar tool in \datasetname, but we aim for a larger-scale dataset. 
Our dataset contains 71K theorems, much more than the 1,602 theorems in Feit--Thompson, and our theorems come from a broad spectrum of \numberofprojects Coq projects.
Another difference is that we augment our dataset with synthetic proofs extracted from the intermediate steps of human proofs. 
Finally, in terms of tactic generation, we generate complete tactics that can be used to obtain full proofs, whereas they group all tactics into categories and only predict the category, not the specific tactic.
Also, they do not predict the location of the arguments in the tactics. Their method for full proof generation is specific to an algebraic rewrite problem, which has only two possible tactics, each with two integer arguments. Their model predicts one tactic (out of 2) and two integers, and is not directly applicable to our setting.

HOList~\citep{bansal2019holist} is a concurrent work introducing a dataset and learning environment for the HOL Light proof assistant~\citep{harrison1996hol}.
HOList consists of 29K proofs solely from the formalization of the Kepler conjecture~\citep{hales2017formal}, a theorem in discrete geometry, whereas {\datasetname } covers more diverse domains including not only pure mathematics but also computer systems.
Similar to ours, HOList also introduces a model for tactic generation. But unlike ours, their method does not generate tactics in the form of ASTs.

\smallsec{Representing and generating programs}
Our model builds upon prior work that uses deep learning to represent and generate programs.
A key ingredient is using a deep network to embed a program into a vector, by treating the program as a sequence of tokens~\citep{allamanis2016convolutional} or using structured input such as ASTs~\citep{allamanis2016learning,alon2018code2vec} or graphs~\citep{allamanis2017learning}.
We use a TreeLSTM~\citep{tai2015improved} on ASTs to embed the input goal and premises.

Similarly, a program can be generated by a deep network as a sequence of tokens~\citep{hindle2012naturalness}, an AST~\citep{parisotto2016neuro} or a graph~\citep{brockschmidt2018generative}.
We adopt the framework of \citet{yin2017syntactic} to generate tactics in the form of ASTs, conditioned on the goal and premises.
However, in our specific task, we face the unique challenge of synthesizing the tactic arguments, which are subject to various semantic constraints, \textit{e.g.}, the \texttt{H} in ``\texttt{apply H}'' must be a valid premise in the current context.
Unlike the purely syntactic approach of \citet{yin2017syntactic}, our model utilizes the semantic constraints to narrow down the output space.

\begin{figure*}[ht]
\vskip 0.2in
\begin{center}
\vspace{-8mm}
\centerline{\includegraphics[width=2.1\columnwidth]{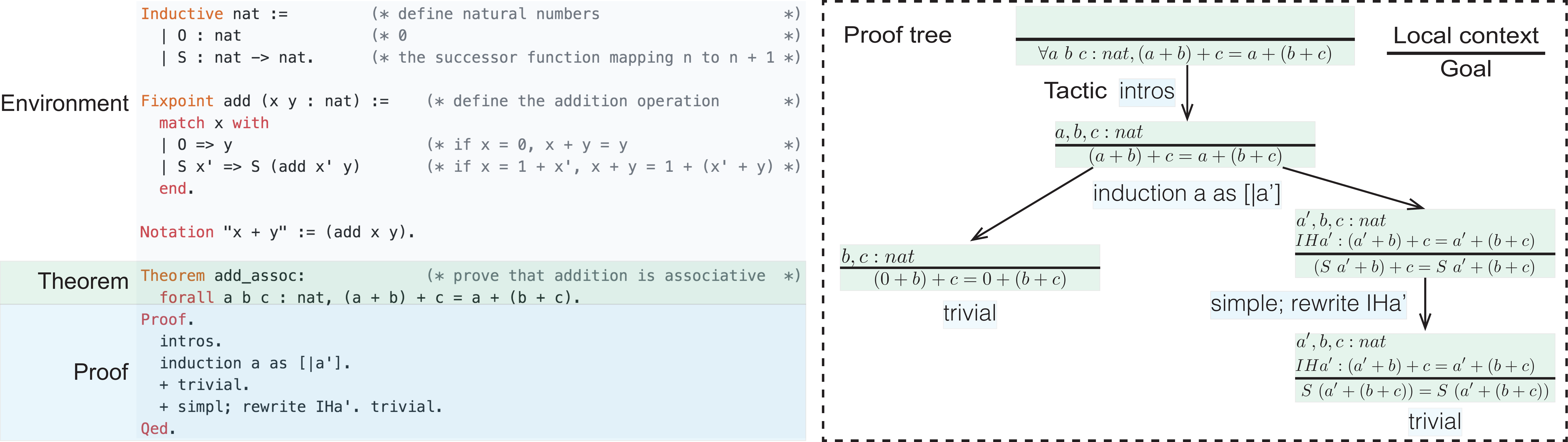}}
\vspace{-3mm}
\caption{\textit{Left}: A simple Coq proof for the associativity of the addition operation on natural numbers. \textit{Right}: The proof tree generated by Coq when executing this proof. A Coq proof consists of a sequences of tactics. We start with the original theorem and apply tactics to decompose the current goal to sub-goals. This process generates a proof tree whose nodes are goals and whose edges are tactics.}
\vspace{-4mm}
\label{fig:example_coq_proof}
\end{center}
\vskip -0.2in
\end{figure*}

\section{Background on Coq}

Coq~\citep{barras1997coq} is a proof assistant with an active community and diverse applications.
It has been used to develop certified software and hardware~\citep{leroy2009formal, paulin1995circuits}, and to prove theorems in mathematics, including the Feit--Thompson theorem~\citep{gonthier2013machine}.
Under the hood of Coq are two pieces of machinery: 
 a functional language for representing mathematical objects, theorems, and proofs, and 
a mechanism for constructing machine-checked proofs semi-automatically.

Coq allows us to define mathematical objects such as sets, functions, and relations. 
For example, we can define the set of natural numbers ($nat$) and the addition operation ($add$) in Fig.~\ref{fig:example_coq_proof} (Left).
These are examples of \textit{terms} in Coq's language.
The runtime \textit{environment} of COq contains a set of current terms, including both user-defined terms and predefined terms from the Coq standard library.
These terms are used to formalize theorems.
As in Fig.~\ref{fig:example_coq_proof}, we state the theorem $\forall a~b~c : nat, (a + b) + c = a + (b + c)$ using $nat$ and $add$.

Theorem proving in Coq is a backward process.
The user starts with the theorem itself as the initial \textit{goal} and repeatedly apply \textit{tactics} to 
decompose the goal into a list of sub-goals (can be an empty list).
The proof is complete when there are no sub-goals left.
Proving is a process of trial and error; the user may try a tactic to decompose the current goal, analyze the feedback from Coq, and backtrack to the previous step to try a different tactic.

A successful Coq proof implicitly generates a \textit{proof tree} whose root is the original theorem and whose nodes are goals (Fig.~\ref{fig:example_coq_proof} \textit{Right}).
All goals share the same environment, but have a unique \textit{local context} with premises local to each goal, such as the induction hypothesis \texttt{IHa'} in Fig.~\ref{fig:example_coq_proof}.

The edges of the proof tree are tactics; they can be simple strings, can have \textit{arguments} at various positions, and can be combined into compound tactics.
For example, \texttt{simpl} simplifies the current goal, ``\texttt{apply H}'' applies a premise \texttt{H}, and ``\texttt{simpl; apply H}'' performs these two operations sequentially.
The space of all valid tactics is described by Ltac, Coq's tactic language~\citep{delahaye2000tactic}.

From a machine learning perspective, theorem proving in Coq resembles a task-oriented dialog~\citep{bordes2016learning}.
The agent interacts with the proof assistant for completing the proof.
At each step, the agent perceives the current goals, their local context, and the global environment;
it then generates an appropriate tactic, which is an expression in Ltac.
Methods for task-oriented dialogues have been based on supervised learning~\citep{bordes2016learning} or reinforcement learning~\citep{liu2017end}.
\datasetname provides human-written proofs as supervision for training dialog agents.
It also allows reinforcement learning on this task when combined with the tool for interacting with Coq.

\section{Constructing \datasetname}

{\datasetname } includes a large-scale dataset of 71K human-written proofs from \numberofprojects open-source software projects in Coq. 
In addition to the source files, we provide abstract syntax trees (ASTs) and rich runtime information of the proofs, including the environments, the goals, and the proof trees.
Furthermore, we propose a novel mechanism for turning intermediate goals into theorems and synthesizing the corresponding proofs.
These synthetic proofs may serve as additional training data.
Further details are in Appendix \ref{appendix:details}.

\smallsec{Processing Coq projects and files}
The source files are organized into projects, which contain a set of inter-related proofs about specific domains. 
The projects in {\datasetname } include the Coq standard library and the packages listed on the Coq Package Index\footnote{\url{https://coq.inria.fr/packages}}.
Some of them may not compile because they require a specific version of Coq, or there is a missing dependency.
We only include the projects that compile.

These projects are split into a training set, a validation set, and a test set.
This ensures that no testing proof comes from a project that is used in training, and makes the dataset suitable for measuring how well the models generalize across various domains.
There are \numberoftrainproofs proofs for training, \numberofvalidproofs proofs for validation and \numberoftestproofs proofs for testing.

We extract the ASTs from the internals of Coq's interpreter.
They are OCaml datatypes.
We serialize them into Lisp-style S-expressions~\citep{mccarthy1960recursive} and provide tools for using them in Python.

\smallsec{Environments, goals, and proof trees}
Proofs are situated in environments containing Coq terms as premises.
We could use the source code to represent the environments, 
as the code completely defines the environment.
However, this is problematic for machine learning models, 
as it burdens them with learning the semantics of Coq code.
Instead, we represent the environments as a collection of \emph{kernel terms}, internal representations used by Coq stripped of syntactic sugar.
This is achieved by executing the proofs and serializing Coq's internals.

In contrast to prior work~\citep{huang2018gamepad}, {\datasetname } supplies the complete environment for each proof---all the premises in the scope, including those defined in the same source file and those imported from other libraries.
Having the complete environment is important because it allows the machine learning model to access all relevant information in structured forms.

We represent each proof as a proof tree, where a node is a goal along with its local context, and an edge is a tactic decomposing the goal into sub-goals.
At each step in a proof, we serialize the current goals from Coq's interpreter and identify the edges in the proof tree by tracking how goals emerge and disappear during the lifetime of the proof.

Environments, goals, and proof trees together form a structured representation of Coq proofs. Compared to raw source code, a structured representation allow machine learning models to more easily exploit the syntactic and semantic structures. It is worth noting that this structured representation is nontrivial to extract because Coq does not provide APIs exposing its internals.
In constructing \datasetname, we modify Coq and use SerAPI~\citep{GallegoArias2016SerAPI} to serialize the runtime information, without touching the core proof-checking module of Coq so as to not compromise the correctness of the proofs.

\smallsec{Synthetic proofs from intermediate goals}
Human-written proofs can be long and complex, making them difficult to learn from. We thus generate shorter proofs from the intermediate goals inside a long proof. We hypothesize that these intermediate goals are easier to prove and more conducive to learning. This also augments the training data with more examples.

For each intermediate goal in a human-written proof, we generate synthetic proofs of length 1, 2, 3, and 4. We detail the generation process in Appendix \ref{appendix:details}.

\smallsec{Dataset statistics}
{\datasetname } has \numberofproofs human-written proofs from \numberofprojects Coq projects.
On average, each proof has 8.7 intermediate goals and 9.1 steps; each step has 10,350.3 premises in the environment and 5.6 premises in the local context; each tactic has 2.0 tokens, and the height of its AST is 1.9.
Among all tactics, 53\% of them contain at least one argument.
Note that these statistics vary significantly across different projects. For example, the average number of premises of a theorem is 13,340 in the CompCert project, but only 661 in the InfSeqExt project.

For the synthetic proofs, we have extracted 159,761 proofs of 1 step; 109,602 proofs of 2 steps; 79,967 proofs of 3 steps and 61,126 proofs of 4 steps.

\section{ASTactic: generating tactics as programs}

\begin{figure*}[ht]
\vskip 0.2in
\begin{center}
\vspace{-8mm}
\centerline{\includegraphics[width=2\columnwidth]{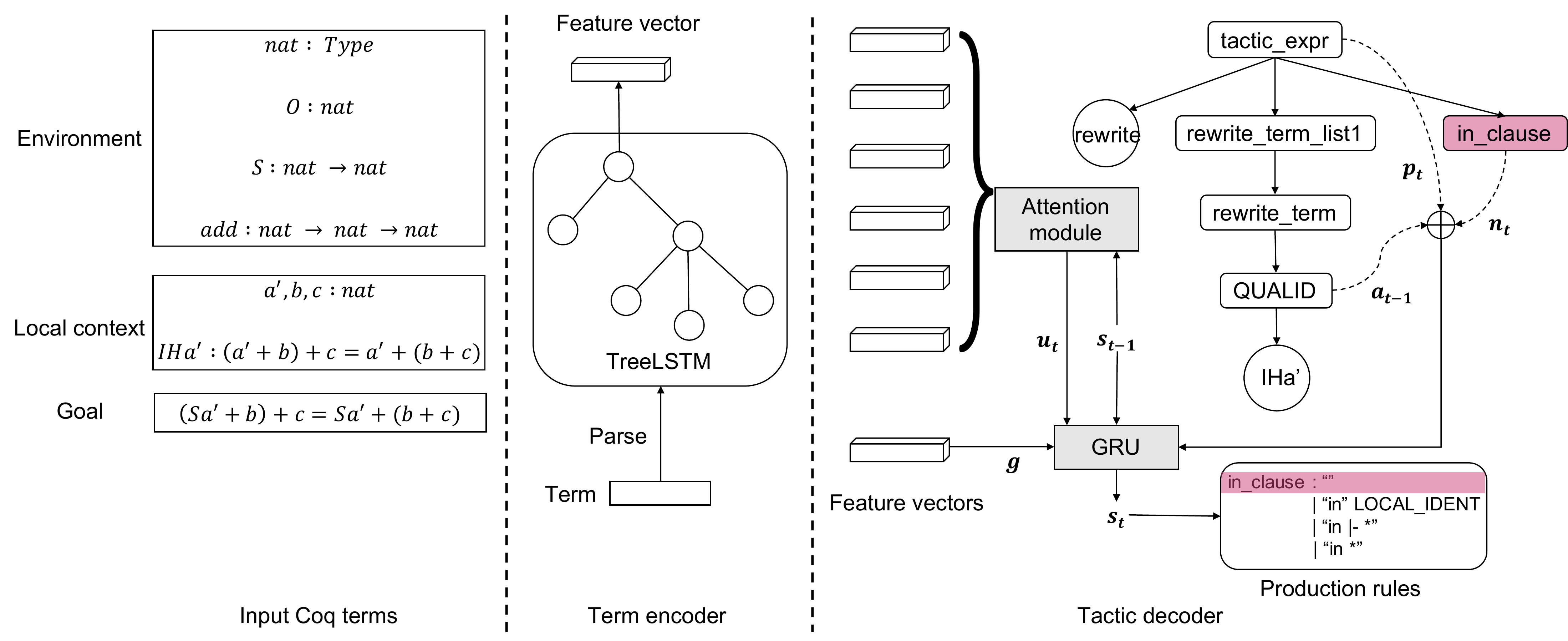}}
\vspace{-4mm}
\caption{The architecture of ASTactic. It generates a tactic AST conditioned on the input Coq terms by sequentially expanding a partial tree. Here we illustrate a single expansion step of the non-terminal node \texttt{in\_clause}. The ASTs of the input terms (\textit{Left}) are encoded into feature vectors by a TreeLSTM network (Middle). A GRU controller then combines them with the information in the partial tree. It updates the decoder state $\mathbf{s_t}$ and uses $\mathbf{s_t}$ to predict the production rule to apply. In this example, the tactic AST is complete (\texttt{rewrite IHa'}) after expanding the current node.}
\vspace{-4mm}
\label{fig:astactic}
\end{center}
\vskip -0.2in
\end{figure*}

We propose a method that proves theorems by interacting with Coq.
At the core of our method is ASTactic---a deep learning model that generates tactics as programs.
Compared to prior work that chooses tactics from a fixed set~\citep{huang2018gamepad,gransden2015sepia,gauthier2018learning,bansal2019holist}, we generate tactics dynamically in the form of abstract syntax trees (ASTs) and synthesize arguments using the available premises during runtime.
At test time, we sample multiple tactics from the model.
They are treated as possible actions to take, and we search for a complete proof via depth-first search (DFS).

\smallsec{Space of tactics}
The output space of our model is specified by a context-free grammar (CFG) that is fixed during training and testing.
Statistics of {\datasetname } show that many valid tactics~\citep{delahaye2000tactic} are seldom used in proofs.
Therefore we simplify the tactic grammar to facilitate learning, at the expense of giving up on some cases. 
Note that these are design choices of the model, not the dataset---We train the model to generates tactics only in the simplified space, but the theorems in the testing data can require tactics outside this space.

We only generate atomic tactics while excluding compound ones such as ``\texttt{tac1; tac2}''.
This is not a severe handicap because all proofs can be completed without compound tactics.
For each type of tactic, we count its number of occurrences in {\datasetname } and manually include the common ones in our tactic space.
When a tactic requires a Coq term as its argument, we constrain the term to be an identifier.
We also exclude user-defined tactics.
The complete CFG is in Appendix \ref{appendix:tac_space}.

\smallsec{Overall model architecture}
ASTactic has an encoder-decoder architecture (Fig.~\ref{fig:astactic}). 
The input and output of the model are both trees.
The encoder embeds all input Coq terms: the goal and the premises expressed in ASTs. 
Conditioned on the embeddings, the decoder generates a program-structured tactic by sequentially growing an AST.

We follow prior works~\citep{tai2015improved, yin2017syntactic} for encoding and decoding trees.
The unique challenge in our task is to synthesize tactic arguments.
In the decoder, we incorporate semantic constraints on the arguments to narrow down the search space.

\smallsec{Encoding the goal and premises}
ASTactic encodes the current goal and premises into vectors.
We include the entire local context and up to 10 premises in the environment, excluding a large number of premises imported from libraries.
A model could be more powerful if it is capable of selecting relevant premises from the entire environment, but that is left for future research.

Both the goal and the premises are Coq terms in the form of AST (Fig.~\ref{fig:astactic} \textit{Left}), and we encode them using a TreeLSTM network~\citep{tai2015improved}. 
Specifically, each node in an AST has a symbol $\mathbf{n}$ indicating its syntactical role.
The network associate each node with a hidden state $\mathbf{h}$ and a memory cell $\mathbf{c}$ which are updated by its children as follows:
\[
(\mathbf{c}, \mathbf{h}) = f_{update}(\mathbf{n}, \mathbf{c_1}, \cdots, \mathbf{c_K}, \sum_{i=1}^{K}{\mathbf{h_i}}),
\]
where the update function $f_{update}$ is the child-sum variant of TreeLSTM, $\mathbf{n}$ is the symbol of the node in one hot encoding, and $\mathbf{c_i}$ and $\mathbf{h_i}$ are the states of the children.

We perform this computation bottom-up and represent the entire tree by $\mathbf{h_{root}}$, the hidden state of the root.
Finally, we append $\mathbf{h_{root}}$ with a 3-dimensional one hot vector;
it indicates whether the term is the goal, a premise in the environment, or a premise in the local context.

\smallsec{Tracking the decoder state}
Conditioned on the input embeddings, the decoder (Fig.~\ref{fig:astactic} \textit{Right}) follows the method in \citet{yin2017syntactic} to generate program-structured tactics as ASTs.
It begins with a single node and grows a partial tree in the depth-first order. At a non-terminal node, it expands the node by choosing a production rule in the CFG of the tactic space. At a terminal node, it emits a token corresponding to a tactic argument.

This sequential generation process is controlled by a gated recurrent unit (GRU)~\cite{cho2014learning}, whose hidden state is updated by the input embeddings and local information in the partially generated AST. 

Formally, we have learnable embeddings for all symbols and production rules in the tactic grammar. At time step $t$, let $\mathbf{n_t}$ be the symbol of the current node;
$\mathbf{a_{t-1}}$ is the production rule used to expand the previous node; $\mathbf{p_t}$ is the parent node's state concatenated with the production rule used to expand the parent;
$\mathbf{g}$ is the goal, which is fixed during the generation process.
The state $\mathbf{s_t}$ is updated by:
\begin{equation}
\mathbf{s_t} = f_{GRU}(\mathbf{s_{t-1}}, [\mathbf{a_{t-1}} : \mathbf{p_t} : \mathbf{n_t}: \mathbf{g}  : \mathbf{u_t}])
\end{equation}
where ``:'' denotes vector concatenation. 
The $\mathbf{u_t}$ above is a weighted sum of premises.
We use $\mathbf{s_{t-1}}$ to compute an attention mask on the premises, which selectively attends to the relevant premises for the current generation step.
The mask is then used to retrieve $\mathbf{u_t}$:
\begin{eqnarray}
w_i = f_{att}(\mathbf{s_{t-1}} : \mathbf{r_i}) \\
\label{eqn:att}
\mathbf{u_t} = \sum_{i}{w_i \mathbf{r_i}}
\end{eqnarray}
where $\mathbf{r_i}$ is the $i$th premise and $w_i$ is its weight. 
$f_{att}$ is a two-layer fully-connected network.

\smallsec{Expanding ASTs and synthesizing arguments}
The state $\mathbf{s_t}$ determines how to expand the tree including which production rules to apply and which tokens to generate.

To select a production rule, we model the probabilities for the rules as:
\begin{equation}
\mathbf{p_t} = \mathrm{softmax}(\mathbf{W_R} \cdot f(\mathbf{s_t})),
\label{eqn:prob}
\end{equation}
where $f$ is a linear layer followed by $tanh$, and 
$\mathbf{W_R}$ is the embedding matrix for production rules.
We expand the node using the applicable rule with the largest probability.

The tokens in the ASTs correspond to the tactic arguments.
Synthesizing them is challenging because the syntactic output space is large:
all valid identifiers in Coq.
However, there are strong semantic constraints on the arguments.
For example, the tactic ``\texttt{apply H}'' applies a premise \texttt{H} to the goal.
The argument \texttt{H} must be a valid premise either in the environment or in the local context.

To leverage the semantic constraints in synthesizing arguments, we group arguments into categories and take different actions for each category.

\begin{itemize}

\item Identifiers of premises (as in ``\texttt{apply H}''): We score each premise using $\mathbf{s_t}$ in the same way as computing the attention masks (Equation.~\ref{eqn:att}).
A softmax on the scores gives us the probability for each premise.

\item Integers (as in ``\texttt{constructor 2}''): Most integers in the data are in the range of [1, 4]. 
We use a 4-way classifier to generate them.

\item Quantified variables in the goal (as in ``\texttt{simple induction n}''): We randomly pick a universally quantified variable in the goal.

\end{itemize}

\smallsec{Training and inference}
We train the model on the proof steps extracted from \datasetname.
When expanding a node using a production rule, we apply the cross-entropy loss to maximize the likelihood of the ground truth production rule in Equation.~\ref{eqn:prob}.
However, when the model emits a tactic argument, there may be no corresponding argument in the ground truth;
because the model might have generated a different tactic from the ground truth.
For example, the model may output ``\texttt{apply H}'' with an argument H, while the ground truth may be \texttt{split} without any argument.

To apply a reasonable loss in this scenario, we train the model with teacher forcing~\cite{williams1989learning}.
During the sequential generation of a tactic AST, the model outputs how to expand the partial tree, but the tree grows following the ground truth, not the model's output. 
Then the arguments generated by the model must correspond to those in the ground truth,
and we can apply losses normally.

During testing, we combine the model with depth-first search (DFS) for fully-automated theorem proving.
At each step, the model samples a few tactics via beam search, which are used to search for a complete proof via DFS. We prune the search space by backtracking when a duplicate proof state is detected.

\smallsec{Implementation details}
We use 256-dimensional vectors for all embeddings in ASTactic, including Coq terms ($\mathbf{g}$, $\mathbf{r_i}$),  production rules ($\mathbf{a_{t-1}}$), GRU hidden state ($\mathbf{s_t}$), and symbols in the tactic grammar ($\mathbf{n_t}$).
The training data includes \numberofhumantrainvalidsteps steps from human-written proofs. 
We do not train ASTactic with synthetic proofs since they only contain tactics extracted from the human proofs. For a method to benefit from synthetic proofs, it should model the entire sequence of tactics rather than an individual tactic.

We train the model using RMSProp~\citep{Tieleman2012} with a learning rate of $3 \times 10^{-5}$ and weight decay of $10^{-6}$.
The training goes for 5 epochs, which takes a few days on a single GeForce GTX 1080 GPU. During testing, our system performs beam search with a beam width of 20 to generate the top 20 tactics at each proof step. And we set a depth limit of 50 during DFS.

\section{Experiments}

\vspace{2mm}

\smallsec{Experimental setup} We evaluate ASTactic on the task of fully-automated theorem proving in Coq, using the \numberoftestproofs testing theorems in \datasetname.
The agent perceives the current goals, their local context, and the environment.
It interacts with Coq by executing commands, which include tactics, backtracking to the previous step (\texttt{Undo}), and any other valid Coq command.
The goal for the agent is to find a complete proof using at most 300 tactics and within a wall time of 10 minutes.
We run all testing experiments on machines with 16GB RAM and two Intel® Xeon® Silver 4114 CPU Cores. 
We do not use GPUs for testing since the speed bottleneck is executing tactics rather than generating tactics.

We compare the performance of our system with several baselines. Our first set of baselines are Coq's built-in automated tactics, including \texttt{trivial}, \texttt{auto}, \texttt{intuition}, and \texttt{easy}. 
They all try to prove the theorem via some simple procedures such as backward chaining.
The second baseline is \texttt{hammer}~\citep{czajka2018hammer}---a hammer-based system that proves theorems using external ATP systems.
In our particular configuration, \texttt{hammer} simultaneously invokes Z3~\cite{de2008z3}, CVC4~\citep{BCD+11}, Vampire~\citep{kovacs2013first}, and E Prover~\citep{schulz2013system}, and returns a proof as long as one of them succeeds.

If we treat \texttt{hammer} as a black box tactic, it sets a default time limit of 20 seconds to the external ATP systems. 
We test \texttt{hammer} both in this setting and in a setting where we extend the time limit to 10 minutes.

All of these automated tactics either prove the goal completely or leave the goal unchanged; they do not decompose the goal into sub-goals.
We combine ASTactic with them as follows: 
At each step, the agent first uses an automated tactic (\textit{e.g.} \texttt{hammer}) to see if it can solve the current goal.
If not, the agent executes a tactic from ASTactic to decompose the goal into sub-goals.

\begin{table}[ht]
\vspace{-5mm}
\caption{Percentage of theorems successfully proved. Our method significantly outperforms Coq's built-in automated tactics. It achieves the highest success rate when combined with  \texttt{hammer}. The default time limit for \texttt{hammer} is 20 seconds and the extended time limit is 10 minutes.} 
\vspace{-2mm}
\label{table:results}
\vskip 0.15in
\begin{center}
\begin{small}
\begin{tabular}{lcc}
\toprule
Method & Success rate (\%) \\
\midrule
trivial & 2.4  \\
auto & 2.9 \\
intuition & 4.4  \\
easy & 4.9  \\
hammer (default time limit) & 17.8  \\
hammer (extended time limit) & 24.8  \\
\midrule
ours & 12.2  \\
ours + auto & 12.8  \\
ours + hammer & \textbf{30.0} \\
\bottomrule
\end{tabular}
\end{small}
\end{center}
\vskip -0.2in
\end{table}

\smallsec{Success rates}
Table~\ref{table:results} shows the percentage of theorems successfully proved. 
Our system proves 12.2\% of the theorems, while the built-in automated tactics in Coq prove less than 4.9\%. While our model underperforms \texttt{hammer} (24.8\% with extended time limit), its performance is nonetheless very encouraging considering that \texttt{hammer} invokes four state-of-the-art external ATP systems that took many years to engineer whereas our model is trained from scratch with a very limited amount of hand engineering. 
When combined with \texttt{hammer}, ASTactic can prove 30.0\% of the theorems, a large improvement of 5.2\% over using \texttt{hammer} alone. This demonstrates that our system can generate effective tactics and can be used to prove theorems previously not provable by automatic methods.

\begin{table}[ht]
\vspace{-4mm}
\caption{The effect of the beam width on the success rate and the average runtime for proving a theorem.}
\vspace{-2mm}
\vspace{-2mm}
\label{table:beam_width}
\begin{center}
\begin{small}
\resizebox{1.0\columnwidth}{!}{
\begin{tabular}{lccccccc}
\toprule
Beam width & 1 & 5 & 10 & 15 & 20 & 25  \\
\midrule
Success rate (\%) & 1.0 & 6.5 & 10.8 & 12.0 & 12.2 & 11.7  \\
Average runtime (seconds) & 0.2 & 1.2 & 2.2 & 2.7 & 3.3  & 3.9  \\
\bottomrule
\end{tabular}
}
\end{small}
\end{center}
\vskip -0.1in
\end{table}

\smallsec{Efficiency}
The beam width is the number of candidate tactics to explore at each proof step; it thus controls the trade-off between speed and accuracy.
Table~\ref{table:beam_width} shows how the beam width affects the runtime and the number of proved theorems.
A large beam width increases the success rate as it enables the model to explore larger search space at the expense of longer runtime. However, when the beam width goes above 20, the success rate drops, probably due to the model being trapped in an unpromising branch for too long.

\begin{figure}[ht]
    \vspace{-5mm}
    \centering
    \subfloat{{\includegraphics[width=0.53\columnwidth]{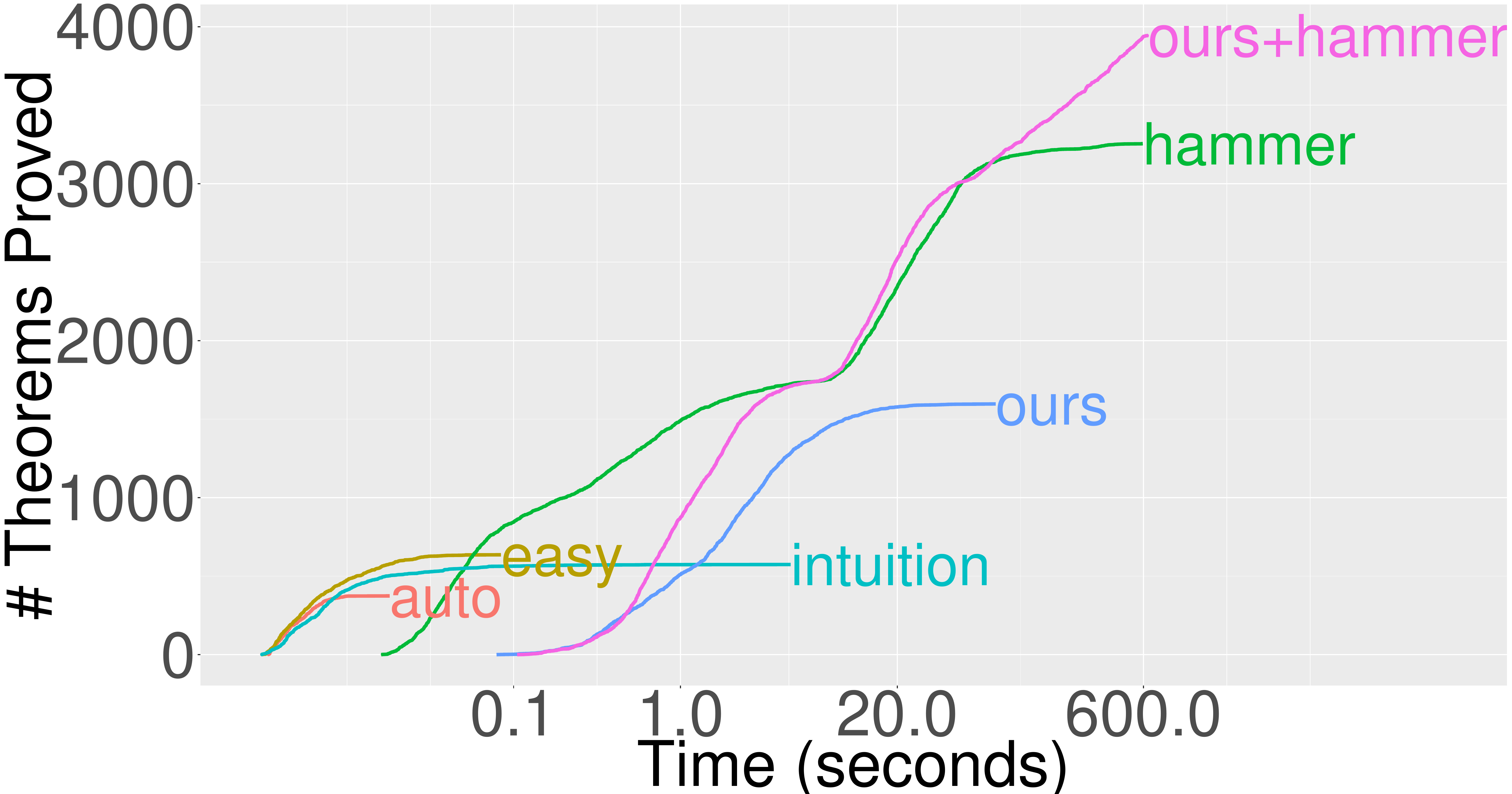} }}
    \subfloat{{\includegraphics[width=0.47\columnwidth]{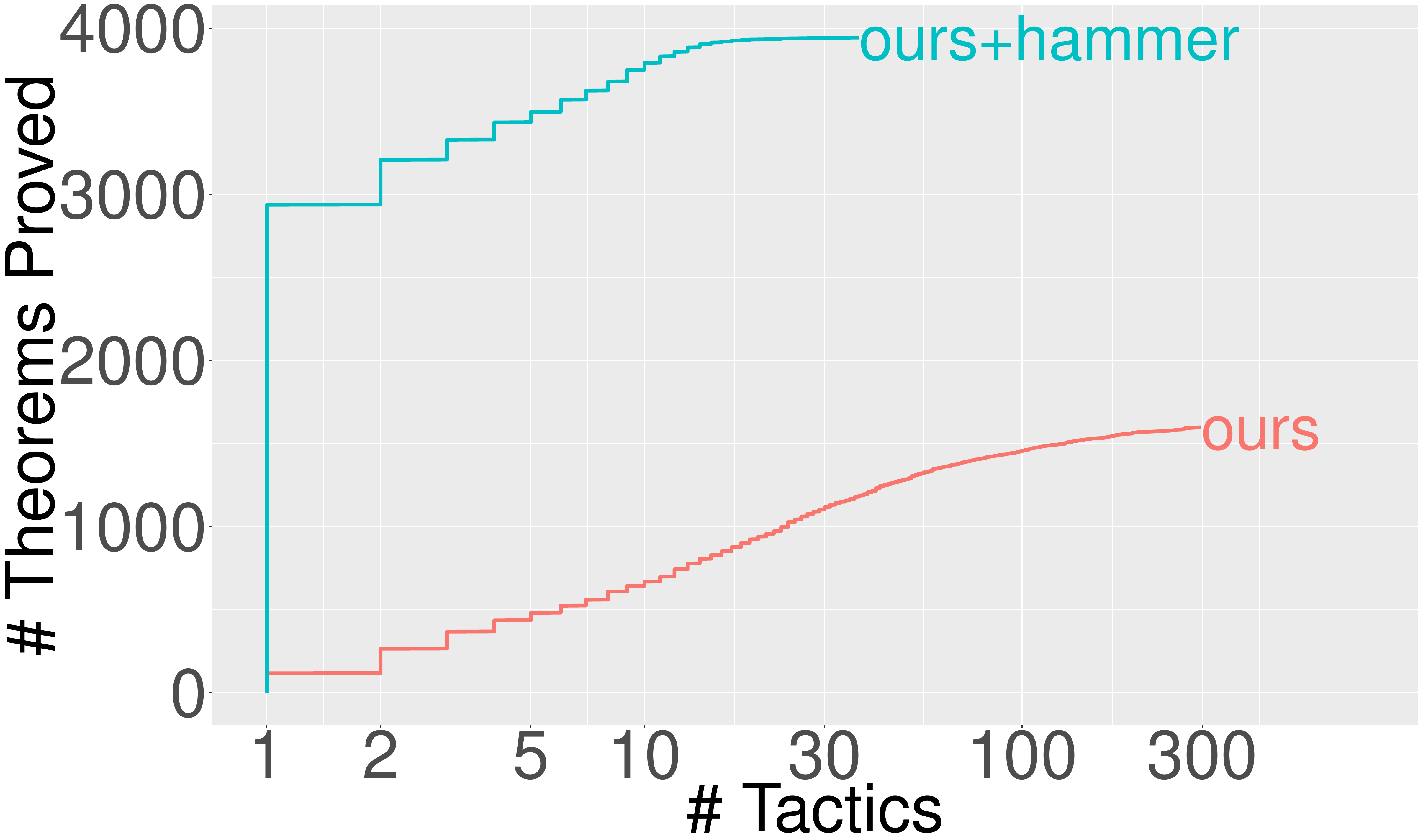}}}
    \vspace{-2mm}
    \caption{The number of proved theorems increases with the runtime (\textit{Left}) and the number of allowed tactics (\textit{Right}).}
    \label{fig:runtime}
\end{figure}

Fig.~\ref{fig:runtime} (\textit{Left}) illustrates the runtime of various methods. The built-in automated tactics are faster, but they can prove only a few hundred theorems. ASTactic combined with \texttt{hammer} takes less than 100 seconds per theorem for proving over 3,000 theorems.

Fig.~\ref{fig:runtime} (\textit{Right}) shows the number of tactics tried before successfully finding a proof.
Compared to SEPIA~\citep{gransden2015sepia}, which uses 10,000 tactics, 
ASTactic is more efficient in exploring the tactic space, needing only a few hundred tactics. When combined with \texttt{hammer}, it typically finds a proof within 10 tactics.

\begin{figure}[ht]
\vskip 0.2in
\begin{center}
\vspace{-7mm}
\centerline{\includegraphics[width=1.0 \columnwidth]{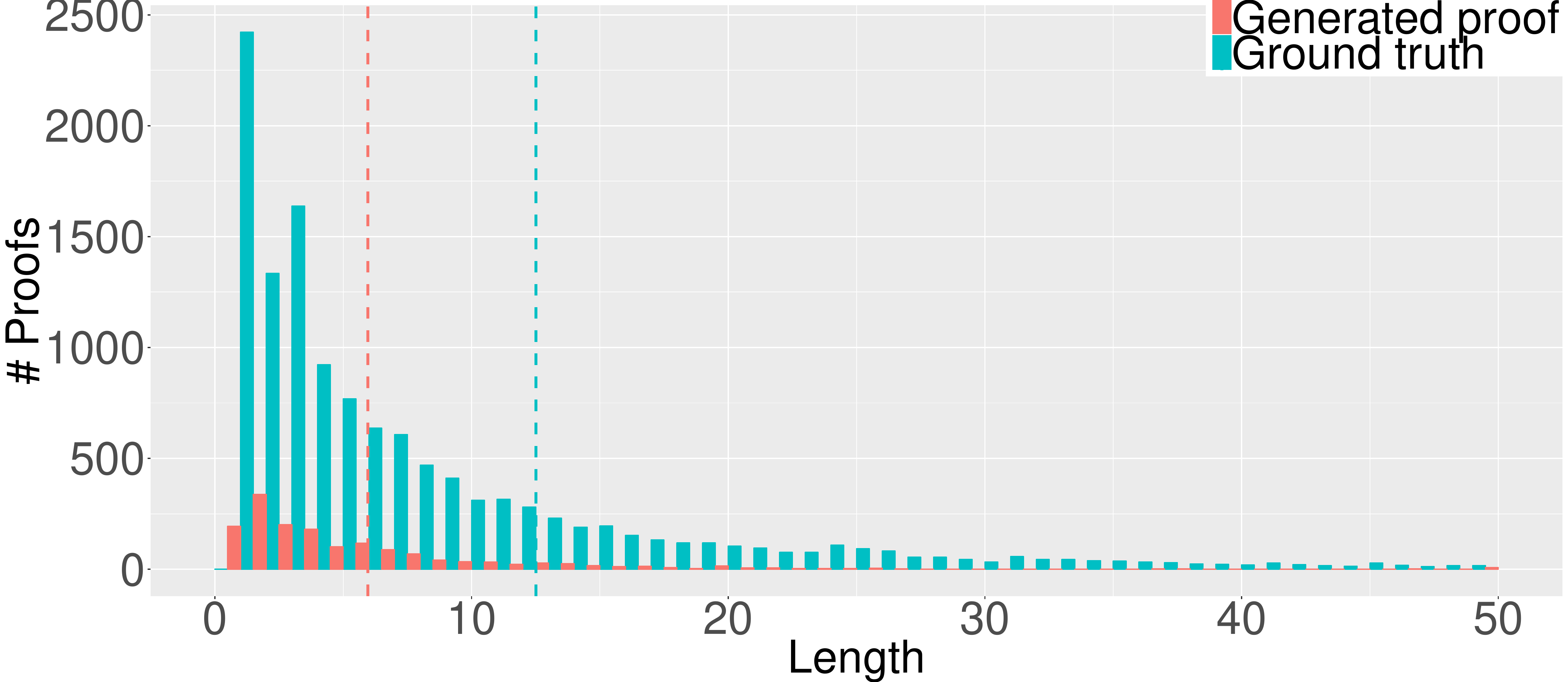}}
\vspace{-3mm}
\caption{The lengths (the number of steps) of the generated proofs compared to all proofs in the testing set. The dash lines are the average lengths (\textit{6.0} and \textit{12.5} respectively).}
\label{fig:length}
\end{center}
\vskip -0.3in
\end{figure}

\smallsec{Generated proofs} Fig.~\ref{fig:length} shows the lengths of the generated proofs compared to all ground truth proofs in the testing set.
As expected, most generated proofs are short (less than 10 steps).
The average length of the generated proofs is 6.0 while the average of the entire testing set is 12.5, which suggests that theorems with longer proofs are much more challenging for the model.

\begin{figure}[h]
\vskip 0.2in
\begin{center}
\vspace{-7mm}
\centerline{\includegraphics[width=\columnwidth]{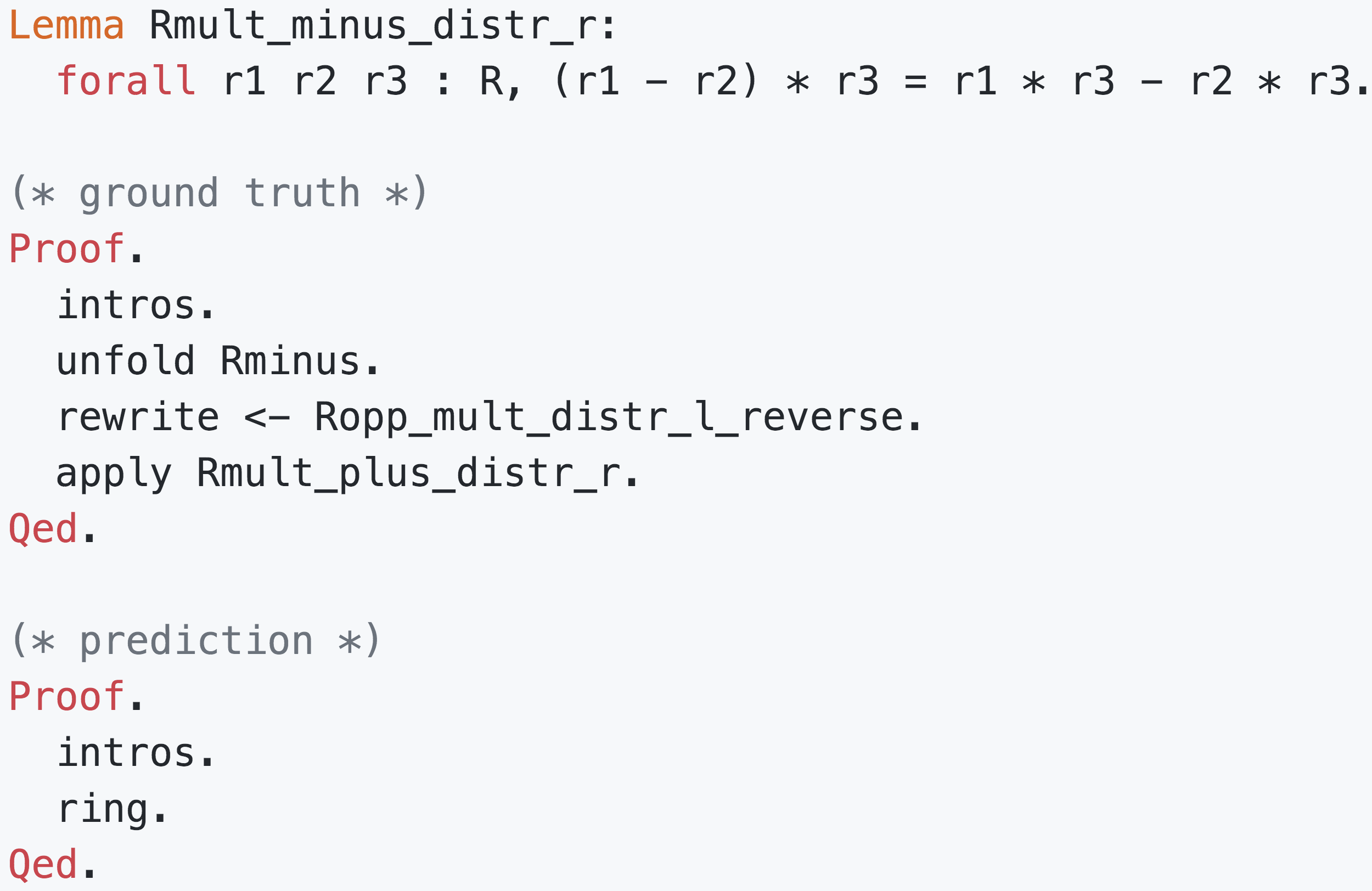}}
\vspace{-2mm}
\caption{An example proof generated by our method. It is shorter than the ground truth thanks to the \texttt{ring} decision procedure.}
\label{fig:generated_proofs}
\end{center}
\vskip -0.3in
\end{figure}

Even for the same theorem, a generated proof can be much shorter than the ground truth. Fig.~\ref{fig:generated_proofs} is one such example. The generated proof calls a decision procedure (\texttt{ring}) to solve the goal with fewer tactics.
This also reflects the fact that longer proofs are harder to find.

\section{Conclusion}

We address the problem of learning to prove theorems in Coq. We have constructed \datasetname---a large-scale dataset and learning environment with human-written proofs from a broad spectrum of Coq projects.
We have developed ASTactic, a deep-learning based model that generates Coq tactics in the form of AST.
Experimental results on {\datasetname } confirm the effectiveness of our model for synthesizing complete proofs automatically.

\section*{Acknowledgements}

This work is partially supported by the National Science Foundation under
Grant No. 1633157.

\renewcommand\thefigure{\Alph{figure}}
\begin{figure*}[ht]
\begin{center}
\centerline{\includegraphics[width=2.0\columnwidth]{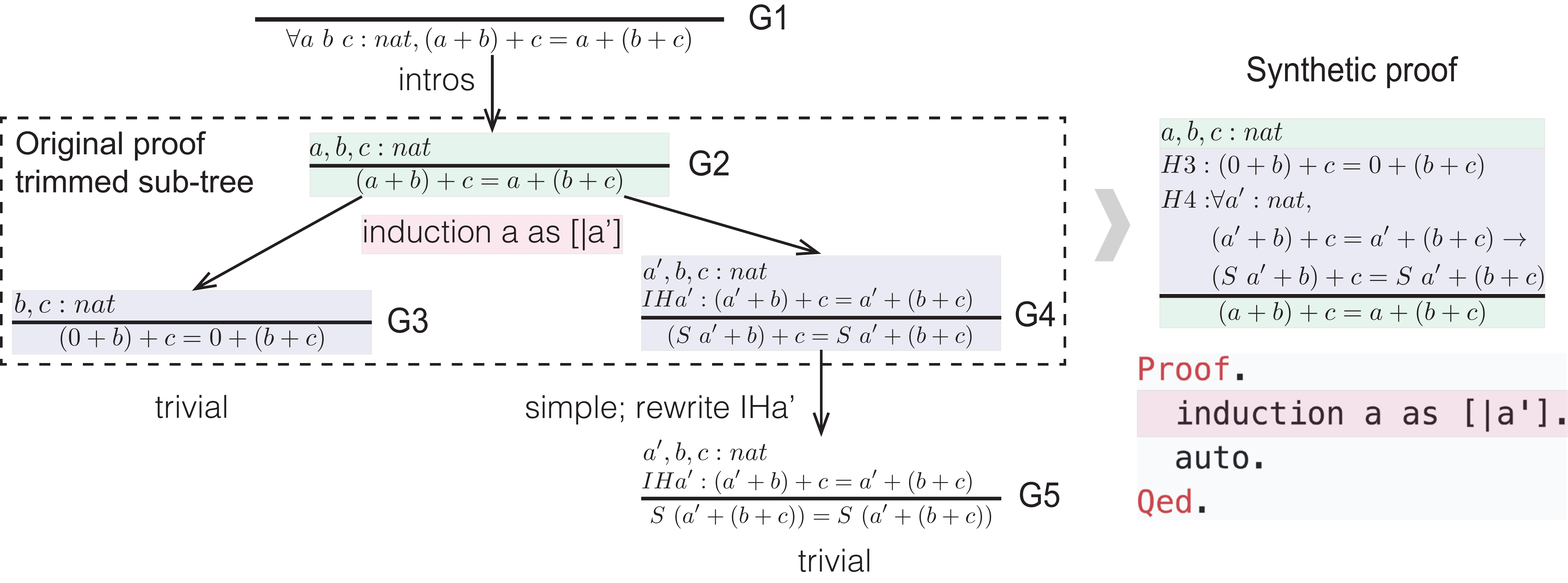}}
\caption{Extracting a synthetic proof from the intermediate goal G2. Goals G3 and G4 are converted into premises in G2's local context. The synthetic proof corresponds to a trimmed sub-tree rooted at G2.}
\label{fig:subproof}
\end{center}
\end{figure*}

\renewcommand{\appendixpagename}{Appendix}
\appendix
\appendixpage

\section{Details on Constructing the Dataset}
\label{appendix:details}

\subsection{Building the Coq projects}

We manually compile and install the Coq standard library and a few projects (such as math-comp) that are frequently required by other projects.
For the rest, we try compiling them automatically using simple commands such as  ``\texttt{./configure \&\& make}'', and we take whatever compiles,
ending up with \numberofprojects projects and \numberoffiles Coq files (excluding the files that do not contain any proof).

\subsection{Reconstructing the Proof Tree}

After applying a tactic, the current goal disappears, and a set of new goals emerge, which become the children of the current goal in the proof tree. 
We can identify the edges of the tree by tracking how goals emerge during the proof.
For example, if the list of goals changes from [2, 7] to [8, 9, 7], we know that node 2 has two children: 8 and 9. 

In certain cases, a tactic can affect more than one goal, and it is unclear who should be the parent node. This can happen when a tactic is applied to multiple goals using a language feature called goal selectors (by default, a tactic is applied only to the first goal). 
However, goal selectors are rarely used in practice.
We discard all such proofs and lose only less than 1\% of our data.
For the remaining data, only one goal disappears at each step, and we can build the proof trees unambiguously.

\subsection{Extracting Synthetic Proofs from Intermediate Goals}

Given an intermediate goal, it is straightforward to treat it as a theorem by adding its local context to the environment.
For example, in Fig.~\ref{fig:subproof}, the goal G2 can be a theorem $(a + b) + c = a + (b + c)$ in the environment augmented by $a$ $b$ and $c$.
Extracting synthetic proofs for the new theorem requires nontrivial processing.
One straightforward proof would be the sequence of tactics that follows G2 in the original human-written proof: ``\texttt{induction a as [$\vert$a']. trivial. simpl; rewrite IHa'. trivial.}''. This proof corresponds to the sub-tree rooted at G2.

However, there are potentially shorter proofs for G2 using a trimmed sub-tree.
For example, if we only apply the first tactic to generates G3 and G4,
then we can treat them as premises H3 and H4, and complete the proof by ``\texttt{apply H3. apply H4.}''. Equivalently, we can also use \texttt{auto} to complete the proof.
This technique of converting unsolved sub-goals into premises allows us to generate synthetic proofs of controllable lengths, by taking a sequence of tactics from the original proof and appending an \texttt{auto} at the end.

We need to take extra care in converting a goal into a premise. For example, it is easy to treat G3 as a premise, but G4 needs some care.
G4 depends on \texttt{a'}, which is missing in G2's context.
In order to convert G4 into a well-formed term in G2's context,
we apply the ``\texttt{generalize dependent}'' tactic to push a local premise into the statement of the goal.
When applied to G4, it generates H4 in Fig.~\ref{fig:subproof}, which can be added to G2's local context.

\clearpage
\onecolumn

\section{The Space of Tactics for ASTactic}
\label{appendix:tac_space}

Below is the context-free grammar in extended Backus-Naur form for the tactic space.
The start symbol is $tactic\_expr$.

\begin{lstlisting}[mathescape]
tactic_expr : intro
            | `apply' term_commalist1 reduced_in_clause
            | `auto' using_clause with_hint_dbs
            | `rewrite' rewrite_term_list1 in_clause
            | `simpl' in_clause
            | `unfold' qualid_list1 in_clause
            | destruct
            | induction
            | `elim' QUALID
            | `split'
            | `assumption'
            | trivial
            | `reflexivity'
            | `case' QUALID
            | clear
            | `subst' local_ident_list
            | `generalize' term_list1
            | `exists' LOCAL_IDENT
            | `red' in_clause
            | `omega'
            | discriminate
            | inversion
            | simple_induction
            | constructor
            | `congruence'
            | `left'
            | `right'
            | `ring'
            | `symmetry'
            | `f_equal'
            | `tauto'
            | `revert' local_ident_list1
            | `specialize' `(' LOCAL_IDENT QUALID `)'
            | `idtac'
            | `hnf' in_clause
            | inversion_clear
            | contradiction
            | `injection' LOCAL_IDENT
            | `exfalso'
            | `cbv'
            | `contradict' LOCAL_IDENT
            | `lia'
            | `field'
            | `easy'
            | `cbn'
            | `exact' QUALID
            | `intuition'
            | `eauto' using_clause with_hint_dbs

LOCAL_IDENT : /[A-Za-z_][A-Za-z0-9_']*/

QUANTIFIED_IDENT : /[A-Za-z_][A-Za-z0-9_']*/

INT : /1|2|3|4/

QUALID : /([A-Za-z_][A-Za-z0-9_']*\.)*[A-Za-z_][A-Za-z0-9_']*/

HINT_DB : /arith|zarith|algebra|real|sets|core|bool|datatypes|coc|set|zfc/

local_ident_list :
                 | LOCAL_IDENT local_ident_list

local_ident_list1 : LOCAL_IDENT
                  | LOCAL_IDENT local_ident_list1

qualid_list1 : QUALID
             | QUALID `,' qualid_list1

term_list1 : QUALID
           | QUALID term_list1

term_commalist1 : QUALID
                | QUALID `,' term_commalist1

hint_db_list1 : HINT_DB
               | HINT_DB hint_db_list1

reduced_in_clause :
                  | `in' LOCAL_IDENT

in_clause :
          | `in' LOCAL_IDENT
          | `in' `|- *'
          | `in' `*'

at_clause : 
          | `at' INT

using_clause :
             | `using' qualid_list1

with_hint_dbs :
              | `with' hint_db_list1
              | `with' `*'

intro : `intro'
      | `intros'

rewrite_term : QUALID
             | `$\rightarrow$' QUALID
             | `$\leftarrow$' QUALID

rewrite_term_list1 : rewrite_term
                   | rewrite_term `,' rewrite_term_list1

destruct : `destruct' term_commalist1

induction : `induction' LOCAL_IDENT
          | `induction' INT

trivial : `trivial' 

clear : `clear'
      | `clear' local_ident_list1

discriminate : `discriminate'
             | `discriminate' LOCAL_IDENT

inversion : `inversion' LOCAL_IDENT
          | `inversion' INT

simple_induction : `simple induction' QUANTIFIED_IDENT
                 |  `simple induction' INT

constructor : `constructor'
            | `constructor' INT

inversion_clear : `inversion_clear' LOCAL_IDENT
                | `inversion_clear' INT

contradiction : `contradiction'
              | `contradiction' LOCAL_IDENT
\end{lstlisting}

\twocolumn

\bibliography{example_paper}

\begin{thebibliography}{51}
\providecommand{\natexlab}[1]{#1}
\providecommand{\url}[1]{\texttt{#1}}
\expandafter\ifx\csname urlstyle\endcsname\relax
  \providecommand{\doi}[1]{doi: #1}\else
  \providecommand{\doi}{doi: \begingroup \urlstyle{rm}\Url}\fi

\bibitem[Allamanis et~al.(2016{\natexlab{a}})Allamanis, Chanthirasegaran,
  Kohli, and Sutton]{allamanis2016learning}
Allamanis, M., Chanthirasegaran, P., Kohli, P., and Sutton, C.
\newblock Learning continuous semantic representations of symbolic expressions.
\newblock \emph{arXiv preprint arXiv:1611.01423}, 2016{\natexlab{a}}.

\bibitem[Allamanis et~al.(2016{\natexlab{b}})Allamanis, Peng, and
  Sutton]{allamanis2016convolutional}
Allamanis, M., Peng, H., and Sutton, C.
\newblock A convolutional attention network for extreme summarization of source
  code.
\newblock In \emph{International Conference on Machine Learning}, pp.\
  2091--2100, 2016{\natexlab{b}}.

\bibitem[Allamanis et~al.(2017)Allamanis, Brockschmidt, and
  Khademi]{allamanis2017learning}
Allamanis, M., Brockschmidt, M., and Khademi, M.
\newblock Learning to represent programs with graphs.
\newblock \emph{arXiv preprint arXiv:1711.00740}, 2017.

\bibitem[Alon et~al.(2018)Alon, Zilberstein, Levy, and Yahav]{alon2018code2vec}
Alon, U., Zilberstein, M., Levy, O., and Yahav, E.
\newblock code2vec: Learning distributed representations of code.
\newblock \emph{arXiv preprint arXiv:1803.09473}, 2018.

\bibitem[Balunovic et~al.(2018)Balunovic, Bielik, and
  Vechev]{balunovic2018learning}
Balunovic, M., Bielik, P., and Vechev, M.
\newblock Learning to solve smt formulas.
\newblock In \emph{Advances in Neural Information Processing Systems}, pp.\
  10317--10328, 2018.

\bibitem[Bancerek et~al.(2015)Bancerek, Byli{\'n}ski, Grabowski,
  Korni{\l}owicz, Matuszewski, Naumowicz, Pak, and Urban]{bancerek2015mizar}
Bancerek, G., Byli{\'n}ski, C., Grabowski, A., Korni{\l}owicz, A., Matuszewski,
  R., Naumowicz, A., Pak, K., and Urban, J.
\newblock Mizar: State-of-the-art and beyond.
\newblock In \emph{Conferences on Intelligent Computer Mathematics}, pp.\
  261--279. Springer, 2015.

\bibitem[Bansal et~al.(2019)Bansal, Loos, Rabe, Szegedy, and
  Wilcox]{bansal2019holist}
Bansal, K., Loos, S.~M., Rabe, M.~N., Szegedy, C., and Wilcox, S.
\newblock Holist: An environment for machine learning of higher-order theorem
  proving (extended version).
\newblock \emph{arXiv preprint arXiv:1904.03241}, 2019.

\bibitem[Barras et~al.(1997)Barras, Boutin, Cornes, Courant, Filliatre,
  Gimenez, Herbelin, Huet, Munoz, Murthy, et~al.]{barras1997coq}
Barras, B., Boutin, S., Cornes, C., Courant, J., Filliatre, J.-C., Gimenez, E.,
  Herbelin, H., Huet, G., Munoz, C., Murthy, C., et~al.
\newblock \emph{The Coq proof assistant reference manual: Version 6.1}.
\newblock PhD thesis, Inria, 1997.

\bibitem[Barrett et~al.(2011)Barrett, Conway, Deters, Hadarean, Jovanovi{'{c}},
  King, Reynolds, and Tinelli]{BCD+11}
Barrett, C., Conway, C.~L., Deters, M., Hadarean, L., Jovanovi{'{c}}, D., King,
  T., Reynolds, A., and Tinelli, C.
\newblock {CVC4}.
\newblock In Gopalakrishnan, G. and Qadeer, S. (eds.), \emph{Proceedings of the
  23rd International Conference on Computer Aided Verification (CAV '11)},
  volume 6806 of \emph{Lecture Notes in Computer Science}, pp.\  171--177.
  Springer, July 2011.
\newblock URL \url{http://www.cs.stanford.edu/~barrett/pubs/BCD+11.pdf}.
\newblock Snowbird, Utah.

\bibitem[Bordes et~al.(2016)Bordes, Boureau, and Weston]{bordes2016learning}
Bordes, A., Boureau, Y.-L., and Weston, J.
\newblock Learning end-to-end goal-oriented dialog.
\newblock \emph{arXiv preprint arXiv:1605.07683}, 2016.

\bibitem[Bridge et~al.(2014)Bridge, Holden, and Paulson]{bridge2014machine}
Bridge, J.~P., Holden, S.~B., and Paulson, L.~C.
\newblock Machine learning for first-order theorem proving.
\newblock \emph{Journal of automated reasoning}, 53\penalty0 (2):\penalty0
  141--172, 2014.

\bibitem[Brockschmidt et~al.(2019)Brockschmidt, Allamanis, Gaunt, and
  Polozov]{brockschmidt2018generative}
Brockschmidt, M., Allamanis, M., Gaunt, A.~L., and Polozov, O.
\newblock Generative code modeling with graphs.
\newblock In \emph{International Conference on Learning Representations}, 2019.
\newblock URL \url{https://openreview.net/forum?id=Bke4KsA5FX}.

\bibitem[Cho et~al.(2014)Cho, Van~Merri{\"e}nboer, Gulcehre, Bahdanau,
  Bougares, Schwenk, and Bengio]{cho2014learning}
Cho, K., Van~Merri{\"e}nboer, B., Gulcehre, C., Bahdanau, D., Bougares, F.,
  Schwenk, H., and Bengio, Y.
\newblock Learning phrase representations using rnn encoder-decoder for
  statistical machine translation.
\newblock \emph{arXiv preprint arXiv:1406.1078}, 2014.

\bibitem[Czajka \& Kaliszyk(2018)Czajka and Kaliszyk]{czajka2018hammer}
Czajka, {\L}. and Kaliszyk, C.
\newblock Hammer for coq: Automation for dependent type theory.
\newblock \emph{Journal of Automated Reasoning}, 61\penalty0 (1-4):\penalty0
  423--453, 2018.

\bibitem[Darvas et~al.(2005)Darvas, H{\"a}hnle, and Sands]{darvas2005theorem}
Darvas, {\'A}., H{\"a}hnle, R., and Sands, D.
\newblock A theorem proving approach to analysis of secure information flow.
\newblock In \emph{International Conference on Security in Pervasive
  Computing}, pp.\  193--209. Springer, 2005.

\bibitem[De~Moura \& Bj{\o}rner(2008)De~Moura and Bj{\o}rner]{de2008z3}
De~Moura, L. and Bj{\o}rner, N.
\newblock Z3: An efficient smt solver.
\newblock In \emph{International conference on Tools and Algorithms for the
  Construction and Analysis of Systems}, pp.\  337--340. Springer, 2008.

\bibitem[Delahaye(2000)]{delahaye2000tactic}
Delahaye, D.
\newblock A tactic language for the system coq.
\newblock In \emph{International Conference on Logic for Programming Artificial
  Intelligence and Reasoning}, pp.\  85--95. Springer, 2000.

\bibitem[Dixon \& Fleuriot(2003)Dixon and Fleuriot]{dixon2003isaplanner}
Dixon, L. and Fleuriot, J.
\newblock Isaplanner: A prototype proof planner in isabelle.
\newblock In \emph{International Conference on Automated Deduction}, pp.\
  279--283. Springer, 2003.

\bibitem[Gallego~Arias(2016)]{GallegoArias2016SerAPI}
Gallego~Arias, E.~J.
\newblock {SerAPI: Machine-Friendly, Data-Centric Serialization for Coq}.
\newblock Technical report, MINES ParisTech, October 2016.
\newblock URL
  \url{https://hal-mines-paristech.archives-ouvertes.fr/hal-01384408}.

\bibitem[Gauthier et~al.(2018)Gauthier, Kaliszyk, Urban, Kumar, and
  Norrish]{gauthier2018learning}
Gauthier, T., Kaliszyk, C., Urban, J., Kumar, R., and Norrish, M.
\newblock Learning to prove with tactics.
\newblock \emph{arXiv preprint arXiv:1804.00596}, 2018.

\bibitem[Gonthier et~al.(2013)Gonthier, Asperti, Avigad, Bertot, Cohen,
  Garillot, Le~Roux, Mahboubi, O’Connor, Biha, et~al.]{gonthier2013machine}
Gonthier, G., Asperti, A., Avigad, J., Bertot, Y., Cohen, C., Garillot, F.,
  Le~Roux, S., Mahboubi, A., O’Connor, R., Biha, S.~O., et~al.
\newblock A machine-checked proof of the odd order theorem.
\newblock In \emph{International Conference on Interactive Theorem Proving},
  pp.\  163--179. Springer, 2013.

\bibitem[Gransden et~al.(2015)Gransden, Walkinshaw, and
  Raman]{gransden2015sepia}
Gransden, T., Walkinshaw, N., and Raman, R.
\newblock Sepia: search for proofs using inferred automata.
\newblock In \emph{International Conference on Automated Deduction}, pp.\
  246--255. Springer, 2015.

\bibitem[Hales et~al.(2017)Hales, Adams, Bauer, Dang, Harrison, Le~Truong,
  Kaliszyk, Magron, McLaughlin, Nguyen, et~al.]{hales2017formal}
Hales, T., Adams, M., Bauer, G., Dang, T.~D., Harrison, J., Le~Truong, H.,
  Kaliszyk, C., Magron, V., McLaughlin, S., Nguyen, T.~T., et~al.
\newblock A formal proof of the kepler conjecture.
\newblock In \emph{Forum of Mathematics, Pi}, volume~5. Cambridge University
  Press, 2017.

\bibitem[Harrison(1996)]{harrison1996hol}
Harrison, J.
\newblock Hol light: A tutorial introduction.
\newblock In \emph{International Conference on Formal Methods in Computer-Aided
  Design}, pp.\  265--269. Springer, 1996.

\bibitem[Harrison et~al.(2014)Harrison, Urban, and
  Wiedijk]{harrison2014history}
Harrison, J., Urban, J., and Wiedijk, F.
\newblock History of interactive theorem proving.
\newblock In \emph{Computational Logic}, volume~9, pp.\  135--214, 2014.

\bibitem[Hindle et~al.(2012)Hindle, Barr, Su, Gabel, and
  Devanbu]{hindle2012naturalness}
Hindle, A., Barr, E.~T., Su, Z., Gabel, M., and Devanbu, P.
\newblock On the naturalness of software.
\newblock In \emph{Software Engineering (ICSE), 2012 34th International
  Conference on}, pp.\  837--847. IEEE, 2012.

\bibitem[Huang et~al.(2019)Huang, Dhariwal, Song, and
  Sutskever]{huang2018gamepad}
Huang, D., Dhariwal, P., Song, D., and Sutskever, I.
\newblock Gamepad: A learning environment for theorem proving.
\newblock In \emph{International Conference on Learning Representations}, 2019.
\newblock URL \url{https://openreview.net/forum?id=r1xwKoR9Y7}.

\bibitem[Irving et~al.(2016)Irving, Szegedy, Alemi, E{\'e}n, Chollet, and
  Urban]{irving2016deepmath}
Irving, G., Szegedy, C., Alemi, A.~A., E{\'e}n, N., Chollet, F., and Urban, J.
\newblock Deepmath-deep sequence models for premise selection.
\newblock In \emph{Advances in Neural Information Processing Systems}, pp.\
  2235--2243, 2016.

\bibitem[Kaliszyk \& Urban(2014)Kaliszyk and Urban]{kaliszyk2014learning}
Kaliszyk, C. and Urban, J.
\newblock Learning-assisted automated reasoning with flyspeck.
\newblock \emph{Journal of Automated Reasoning}, 53\penalty0 (2):\penalty0
  173--213, 2014.

\bibitem[Kaliszyk et~al.(2018)Kaliszyk, Urban, Michalewski, and
  Ol{\v{s}}{\'a}k]{kaliszyk2018reinforcement}
Kaliszyk, C., Urban, J., Michalewski, H., and Ol{\v{s}}{\'a}k, M.
\newblock Reinforcement learning of theorem proving.
\newblock \emph{arXiv preprint arXiv:1805.07563}, 2018.

\bibitem[Kern \& Greenstreet(1999)Kern and Greenstreet]{kern1999formal}
Kern, C. and Greenstreet, M.~R.
\newblock Formal verification in hardware design: a survey.
\newblock \emph{ACM Transactions on Design Automation of Electronic Systems
  (TODAES)}, 4\penalty0 (2):\penalty0 123--193, 1999.

\bibitem[Kov{\'a}cs \& Voronkov(2013)Kov{\'a}cs and Voronkov]{kovacs2013first}
Kov{\'a}cs, L. and Voronkov, A.
\newblock First-order theorem proving and vampire.
\newblock In \emph{International Conference on Computer Aided Verification},
  pp.\  1--35. Springer, 2013.

\bibitem[Leroy(2009)]{leroy2009formal}
Leroy, X.
\newblock Formal verification of a realistic compiler.
\newblock \emph{Communications of the ACM}, 52\penalty0 (7):\penalty0 107--115,
  2009.

\bibitem[Liu et~al.(2017)Liu, Tur, Hakkani-Tur, Shah, and Heck]{liu2017end}
Liu, B., Tur, G., Hakkani-Tur, D., Shah, P., and Heck, L.
\newblock End-to-end optimization of task-oriented dialogue model with deep
  reinforcement learning.
\newblock \emph{arXiv preprint arXiv:1711.10712}, 2017.

\bibitem[Loos et~al.(2017)Loos, Irving, Szegedy, and Kaliszyk]{loos2017deep}
Loos, S., Irving, G., Szegedy, C., and Kaliszyk, C.
\newblock Deep network guided proof search.
\newblock \emph{arXiv preprint arXiv:1701.06972}, 2017.

\bibitem[McCarthy(1960)]{mccarthy1960recursive}
McCarthy, J.
\newblock Recursive functions of symbolic expressions and their computation by
  machine, part i.
\newblock \emph{Communications of the ACM}, 3\penalty0 (4):\penalty0 184--195,
  1960.

\bibitem[McCune(1997)]{mccune1997solution}
McCune, W.
\newblock Solution of the robbins problem.
\newblock \emph{Journal of Automated Reasoning}, 19\penalty0 (3):\penalty0
  263--276, 1997.

\bibitem[Parisotto et~al.(2016)Parisotto, Mohamed, Singh, Li, Zhou, and
  Kohli]{parisotto2016neuro}
Parisotto, E., Mohamed, A.-r., Singh, R., Li, L., Zhou, D., and Kohli, P.
\newblock Neuro-symbolic program synthesis.
\newblock \emph{arXiv preprint arXiv:1611.01855}, 2016.

\bibitem[Paulin-Mohring(1995)]{paulin1995circuits}
Paulin-Mohring, C.
\newblock Circuits as streams in coq: Verification of a sequential multiplier.
\newblock In \emph{International Workshop on Types for Proofs and Programs},
  pp.\  216--230. Springer, 1995.

\bibitem[Paulson(1994)]{paulson1994isabelle}
Paulson, L.~C.
\newblock \emph{Isabelle: A generic theorem prover}, volume 828.
\newblock Springer Science \& Business Media, 1994.

\bibitem[Paulson \& Blanchette(2010)Paulson and Blanchette]{paulson2010three}
Paulson, L.~C. and Blanchette, J.~C.
\newblock Three years of experience with sledgehammer, a practical link between
  automatic and interactive theorem provers.
\newblock In \emph{PAAR@ IJCAR}, pp.\  1--10, 2010.

\bibitem[Rockt{\"a}schel \& Riedel(2017)Rockt{\"a}schel and
  Riedel]{rocktaschel2017end}
Rockt{\"a}schel, T. and Riedel, S.
\newblock End-to-end differentiable proving.
\newblock In \emph{Advances in Neural Information Processing Systems}, pp.\
  3788--3800, 2017.

\bibitem[Schulz(2013)]{schulz2013system}
Schulz, S.
\newblock System description: E 1.8.
\newblock In \emph{International Conference on Logic for Programming Artificial
  Intelligence and Reasoning}, pp.\  735--743. Springer, 2013.

\bibitem[Sutcliffe(2009)]{sutcliffe2009tptp}
Sutcliffe, G.
\newblock The tptp problem library and associated infrastructure.
\newblock \emph{Journal of Automated Reasoning}, 43\penalty0 (4):\penalty0 337,
  2009.

\bibitem[Tai et~al.(2015)Tai, Socher, and Manning]{tai2015improved}
Tai, K.~S., Socher, R., and Manning, C.~D.
\newblock Improved semantic representations from tree-structured long
  short-term memory networks.
\newblock \emph{arXiv preprint arXiv:1503.00075}, 2015.

\bibitem[Tieleman \& Hinton(2012)Tieleman and Hinton]{Tieleman2012}
Tieleman, T. and Hinton, G.
\newblock {Lecture 6.5---RmsProp: Divide the gradient by a running average of
  its recent magnitude}.
\newblock COURSERA: Neural Networks for Machine Learning, 2012.

\bibitem[Urban(2004)]{urban2004mptp}
Urban, J.
\newblock Mptp--motivation, implementation, first experiments.
\newblock \emph{Journal of Automated Reasoning}, 33\penalty0 (3-4):\penalty0
  319--339, 2004.

\bibitem[Urban et~al.(2011)Urban, Vysko{\v{c}}il, and
  {\v{S}}t{\v{e}}p{\'a}nek]{urban2011malecop}
Urban, J., Vysko{\v{c}}il, J., and {\v{S}}t{\v{e}}p{\'a}nek, P.
\newblock Malecop machine learning connection prover.
\newblock In \emph{International Conference on Automated Reasoning with
  Analytic Tableaux and Related Methods}, pp.\  263--277. Springer, 2011.

\bibitem[Wang et~al.(2017)Wang, Tang, Wang, and Deng]{wang2017premise}
Wang, M., Tang, Y., Wang, J., and Deng, J.
\newblock Premise selection for theorem proving by deep graph embedding.
\newblock In \emph{Advances in Neural Information Processing Systems}, pp.\
  2786--2796, 2017.

\bibitem[Williams \& Zipser(1989)Williams and Zipser]{williams1989learning}
Williams, R.~J. and Zipser, D.
\newblock A learning algorithm for continually running fully recurrent neural
  networks.
\newblock \emph{Neural computation}, 1\penalty0 (2):\penalty0 270--280, 1989.

\bibitem[Yin \& Neubig(2017)Yin and Neubig]{yin2017syntactic}
Yin, P. and Neubig, G.
\newblock A syntactic neural model for general-purpose code generation.
\newblock \emph{arXiv preprint arXiv:1704.01696}, 2017.

\end{thebibliography}
\bibliographystyle{icml2019}

\end{document}